\documentclass[sigconf,screen,fleqn]{acmart}

\usepackage{amssymb}
\usepackage[ruled,vlined]{algorithm2e}
\SetKw{KwTo}{to}
\SetKw{KwStep}{step}

\setcopyright{none} 

\acmConference[ICAIF '25]{ACM International Conference on AI in Finance}{2025}{Singapore}

\acmBooktitle{Proceedings of the ACM International Conference on AI in Finance (ICAIF '25)}
\acmYear{2025}
\acmDOI{}
\acmISBN{}

\usepackage[utf8]{inputenc}
\usepackage[T1]{fontenc}

\usepackage{amsmath,amssymb,amsthm,mathtools}
\usepackage{bm}
\usepackage{booktabs}
\usepackage{graphicx}
\usepackage{adjustbox}
\usepackage{tikz}
\usepackage{subcaption}
\usepackage[ruled,vlined]{algorithm2e}
\newcommand{\T}{\mathsf{T}}
\newcommand{\E}{\mathbb{E}}
\usepackage{xcolor}
\usepackage{tabularx, makecell, pifont}

\newcommand{\orgdag}{\textsuperscript{\dag}}     

\newcommand{\corrnote}[1]{%
  \begingroup
  \renewcommand\thefootnote{\fnsymbol{footnote}}
  \footnotetext[1]{Corresponding author: #1}%
  \endgroup
}
\newcommand{\orgdisclaimer}[1]{%
  \begingroup
  \renewcommand\thefootnote{\fnsymbol{footnote}}
  \footnotetext[2]{#1}
  \endgroup
}

\usepackage{tikz}
\usetikzlibrary{arrows.meta,positioning,fit,calc,backgrounds,decorations.pathreplacing,shapes.multipart}
\definecolor{SigPurple}{HTML}{6C5CE7}
\definecolor{RDEBlue}{HTML}{1F77B4}
\definecolor{HeadGreen}{HTML}{2CA02C}
\definecolor{LossRed}{HTML}{D62728}
\definecolor{AuxOrange}{HTML}{FF7F0E}
\definecolor{CanvasBG}{HTML}{F6F7FB}

\tikzset{
  >={Stealth[length=3.1mm,width=2.6mm]},
  thinlink/.style={-Stealth, line width=0.9pt},
  fatlink/.style={-Stealth, line width=1.4pt},
  box/.style={rounded corners=3pt, draw=black!65, fill=white, inner sep=5pt,
              minimum height=10mm, font=\small, align=center},
  sigbox/.style={box, draw=SigPurple!75!black, fill=SigPurple!7},
  rdebox/.style={box, draw=RDEBlue!75!black, fill=RDEBlue!7, minimum width=52mm, minimum height=22mm},
  headbox/.style={box, draw=HeadGreen!75!black, fill=HeadGreen!8, minimum width=42mm},
  lossbox/.style={box, draw=LossRed!85!black, fill=LossRed!6},
  auxbox/.style={box, draw=AuxOrange!85!black, fill=AuxOrange!10},
  legend/.style={draw=black!25, rounded corners=3pt, fill=white, inner sep=4pt, font=\footnotesize}
}

\usepackage{pgfplots}

\usepackage{adjustbox}

\usepackage{tikz}
\usetikzlibrary{arrows.meta,positioning,fit,backgrounds,calc,shapes,decorations.pathmorphing}
\definecolor{oiBlue}{HTML}{0072B2}
\definecolor{oiSky}{HTML}{56B4E9}
\definecolor{oiGreen}{HTML}{009E73}
\definecolor{oiYellow}{HTML}{F0E442}
\definecolor{oiOrange}{HTML}{E69F00}
\definecolor{oiVerm}{HTML}{D55E00}
\definecolor{oiGrey}{HTML}{7F7F7F}

\graphicspath{{figs/}{images/}{.}/}

\pgfplotsset{
  compat=1.18,
  every axis/.append style={
    tick label style={font=\scriptsize},
    label style={font=\scriptsize},
    title style={font=\scriptsize, yshift=-1.2mm},
    legend style={font=\scriptsize, cells={anchor=west}}
  }
}
\usepgfplotslibrary{groupplots,statistics,colorbrewer}

\usetikzlibrary{arrows.meta,positioning,calc,fit,shapes.geometric,shapes.misc,backgrounds,matrix}

\AtBeginDocument{\setlength{\mathindent}{0pt}}

\definecolor{oiBlue}{HTML}{0072B2}
\definecolor{oiGreen}{HTML}{009E73}
\definecolor{oiOrange}{HTML}{E69F00}
\colorlet{laneScenario}{oiBlue!8}
\colorlet{laneSignal}{oiGreen!8}
\colorlet{laneAlloc}{oiOrange!10}

\newcommand{\artifactnote}[1]{%
  \begingroup
  \renewcommand\thefootnote{\fnsymbol{footnote}}
  \footnotetext[3]{#1}%
  \endgroup
}

\begin{document}

\title{Rough Path Signatures: Learning Neural RDEs for Portfolio Optimization}
\author{Ali Atiah Alzahrani\orgdag}
\affiliation{%
  \city{Riyadh}
  \country{Saudi Arabia}
}

\begin{teaserfigure}
  \includegraphics[width=\textwidth]{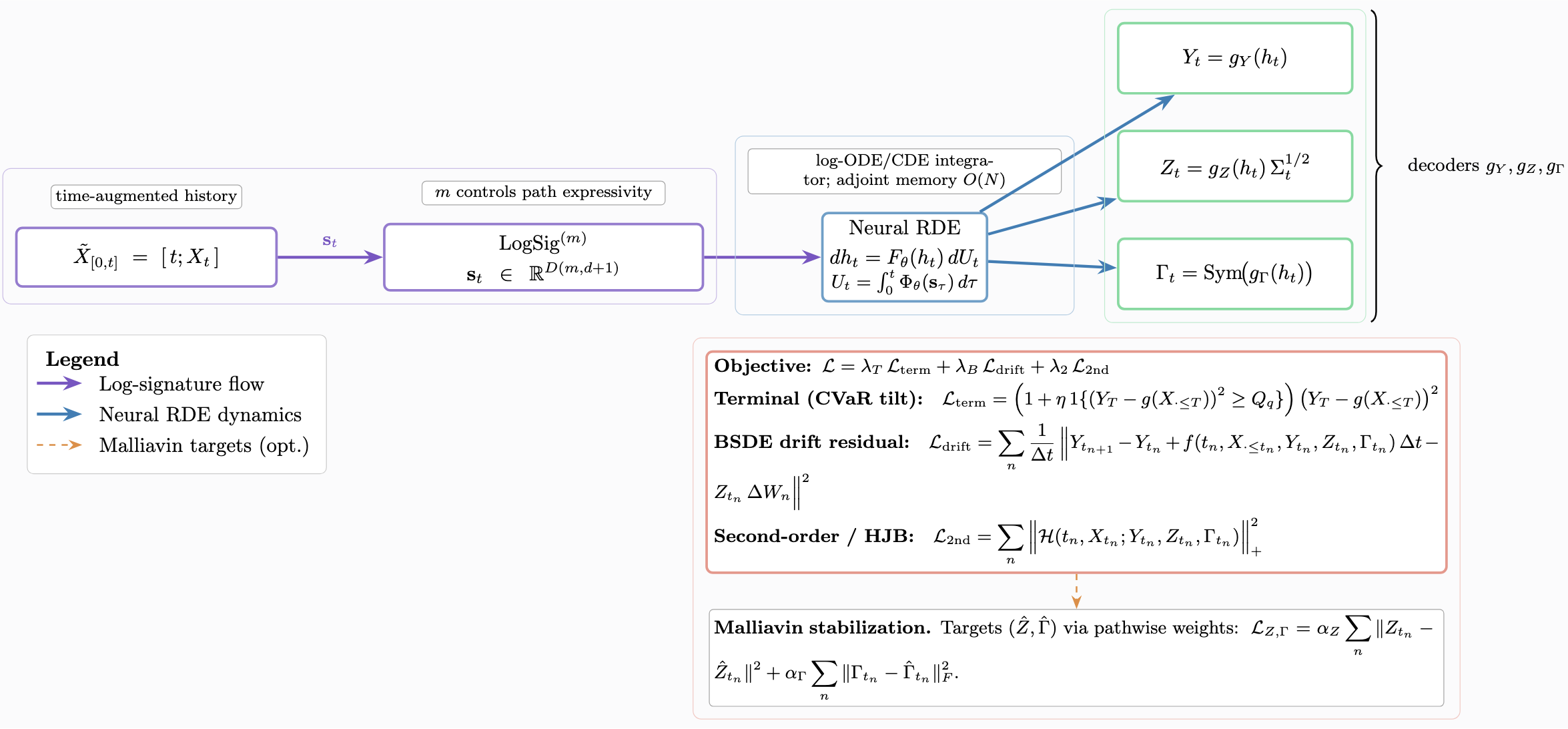}
  \caption{Sig--RDE Architecture for Path-Dependent BSDEs}
  \Description{Enjoying the baseball game from the third-base
  seats. Ichiro Suzuki preparing to bat.}
  \label{fig:teaser}
\end{teaserfigure}

\begin{abstract}
We tackle high-dimensional, path-dependent valuation and control and introduce a deep BSDE/2BSDE solver that couples truncated log-signatures with a neural rough differential equation (RDE) backbone. The architecture aligns stochastic analysis with sequence-to-path learning: a CVaR-tilted terminal objective targets left-tail risk, while an optional second-order (2BSDE) head supplies curvature estimates for risk-sensitive control. Under matched compute and parameter budgets, the method improves accuracy, tail fidelity, and training stability across Asian and barrier option pricing and portfolio control: at \(d=200\) it achieves \(\mathrm{CVaR}_{0.99}=9.80\%\) versus \(12.00\!-\!13.10\%\) for strong baselines, attains the lowest HJB residual \((0.011)\), and yields the lowest RMSEs for \(Z\) and \(\Gamma\). Ablations over truncation depth, local windows, and tilt parameters confirm complementary gains from the sequence-to-path representation and the 2BSDE head. Taken together, the results highlight a bidirectional dialogue between stochastic analysis and modern deep learning: stochastic tools inform representations and objectives, while sequence-to-path models expand the class of solvable financial models at scale.
\end{abstract}



\maketitle

\corrnote{alialzahrani@pif.gov.sa}
\orgdisclaimer{The views expressed are those of the author and do not reflect the views of any other individual or entity. This material is for research purposes only and does not constitute investment advice.}

\artifactnote{Code: \url{https://github.com/AliAtiah/SigRDE}.}

\section{Introduction}
Deep learning has unlocked high-dimensional solvers for semilinear parabolic PDEs via the BSDE representation \cite{Han2018PNAS}, with subsequent advances in local-in-time training (DBDP) \cite{Hure2020DBDP,Hure2020MCOMP}, variance-reduced operator splitting \cite{Beck2019DeepSplitting,Beck2021DeepSplittingSIAM}, and fully nonlinear formulations through 2BSDEs \cite{Beck2017Deep2BSDE,PhamWarin2020FullyNonlinear}. Yet many problems of practical interest in quantitative finance are \emph{path-dependent}—from Asian and barrier features to history-aware portfolio control—so the value functional $u(t,X_{\cdot\le t})$ depends on the entire trajectory. Functional Itô calculus and PPDE theory formalize this non-Markovian setting and its link to BSDEs \cite{Dupire2009,ContFournié2013,Ekren2011Viscosity}, but neural solvers often struggle to encode long-range path information without exploding parameters or memory. We address this gap with a specialized deep BSDE architecture that couples \textbf{truncated log-signature} encoders with a \textbf{Neural Rough Differential Equation (Neural RDE)} backbone for $(Y_t,Z_t)$: signatures provide a principled, hierarchical summary of paths as iterated integrals \cite{Feng2023DeepSignature}, while Neural CDE/RDE models evolve hidden states in continuous time with adjoint memory and stable long-horizon gradients \cite{Morrill2021NRDE,Fang2023NeuralRDEPPDE}. This pairing directly targets the information bottleneck in path dependence, retains Monte-Carlo flexibility, and integrates naturally with finance-specific enhancements: a \emph{risk-sensitive} BSDE objective (CVaR-tilted terminal mismatch) to improve left-tail calibration, and a 2BSDE-compatible second-order head for HJB-type control where we systematically compare \emph{Malliavin} versus \emph{autograd} Hessian estimators under identical time discretizations \cite{Beck2017Deep2BSDE,PhamWarin2020FullyNonlinear}. Our evaluation plan centers on high-dimensional path-dependent pricing (e.g., Asian options) and stochastic-volatility portfolio control, reporting absolute/relative error, runtime, peak memory, and seed-level confidence intervals, alongside ablations on signature depth, RDE vector-field width, multistep depth, and $\Gamma$ estimation. In line with the workshop’s \emph{two-way dialogue}, stochastic analysis informs the design of representations, losses, and estimators, while modern deep sequence-to-path models extend the frontier of solvable financial models to $d\!\gg\!1$ with improved stability and tail fidelity \cite{Han2018PNAS,Hure2020DBDP,Beck2019DeepSplitting,Beck2017Deep2BSDE,Dupire2009,ContFournié2013,Fang2023NeuralRDEPPDE}.

\section{Background and Related Work}
\noindent\textbf{High–dimensional BSDE/PDE solvers.}
Deep neural solvers for semilinear parabolic PDEs via BSDEs have progressed from global end-to-end training \cite{Han2018PNAS} to variance- and stability-focused schemes such as DBDP (local-in-time windows with analysis and error bounds) \cite{Hure2020DBDP,Hure2020MCOMP} and Deep Splitting (operator decomposition) \cite{Beck2019DeepSplitting,Beck2021DeepSplittingSIAM}. In parallel, mesh-free solvers like DGM \cite{SirignanoSpiliopoulos2018DGM} and PINNs \cite{Raissi2019PINNs} introduced residual minimization perspectives that complement BSDE training. For fully nonlinear problems, second-order BSDEs (2BSDEs) provide a probabilistic representation of HJB equations \cite{SonerTouziZhang2012SecondOrderBSDE}; neural instantiations combine BSDE heads with second-order structure and physics penalties \cite{Beck2017Deep2BSDE,PhamWarin2020FullyNonlinear}. Alternative Monte-Carlo lines—regression-based BSDEs \cite{GobetLemorWarin2005BSDERegress,BouchardTouzi2004DiscreteTime} and branching-diffusion representations for semilinear PDEs \cite{HenryLabordere2019Branching}—further illuminate the variance/bias trade-offs that modern deep BSDE pipelines must control.

\noindent\textbf{Path dependence and PPDEs.}
Path dependence arises when the value functional depends on the entire history $X_{\cdot\le t}$. Functional Itô calculus and viscosity solutions for PPDEs formalize this non-Markovian regime and its link to (2)BSDEs \cite{Dupire2009,ContFournié2013,Ekren2011Viscosity}. Neural approaches include PDGM with LSTMs \cite{Saporito2020PDGM}, signature-aware BSDEs \cite{Feng2023DeepSignature}, and Neural RDE solvers tailored to PPDEs \cite{Fang2023NeuralRDEPPDE}. These works point to two complementary levers for high-$d$ path problems: (i) principled summaries of history and (ii) stable continuous-time hidden dynamics.

\noindent\textbf{Signatures and Neural CDE/RDEs.}
The signature of a path (and its log-signature) provides a universal, hierarchical set of iterated integrals with strong approximation properties for path functionals \cite{Lyons1998RoughPaths,ChevyrevKormilitzin2016PrimerSignature}. Efficient software has made high-order signatures practical \cite{ReizensteinGraham2015IISignature}, and deep variants (e.g., deep signature transforms) have shown strong empirical performance on sequential data \cite{Bonnier2019DeepSigTransform}. In parallel, neural differential equations move representation learning into continuous time: Neural ODEs offer adjoint-based memory efficiency \cite{Chen2018NeuralODE}; Neural CDEs/RDEs further align the model class with controlled/rough dynamics and have demonstrated stable long-horizon gradients \cite{KidgerLyons2020NeuralCDE,Morrill2021NRDE}. In finance, combining log-signatures with Neural RDE/CDE backbones is a natural fit for PPDE/BSDE training, enabling controllable path expressivity and continuous-time state propagation \cite{Feng2023DeepSignature,Fang2023NeuralRDEPPDE}.

\noindent\textbf{Estimating $(Z,\Gamma)$ and variance reduction.}
Reliable estimation of the martingale integrand $Z$ and second-order object $\Gamma$ is crucial for both pricing and control. Classical Malliavin-weight identities supply low-variance regression targets for $Z$ and higher-order quantities \cite{FournieEtAl1999MalliavinMC,Glasserman2003MonteCarlo}. In the deep BSDE setting, these targets can be blended with direct heads and drift-residual penalties to stabilize training, particularly for HJB/2BSDEs where second-order information enters the driver \cite{Beck2017Deep2BSDE,PhamWarin2020FullyNonlinear}. Regression-based BSDE estimators \cite{GobetLemorWarin2005BSDERegress,BouchardTouzi2004DiscreteTime} inform practical choices of basis/heads, while local-in-time training (DBDP) reduces variance in long horizons \cite{Hure2020DBDP}.

\noindent\textbf{Risk-sensitive and tail-aware training.}
When left-tail fidelity matters (e.g., barrier features, drawdown-aware control), risk-sensitive objectives help align learning with evaluation. Conditional Value-at-Risk (CVaR) \cite{RockafellarUryasev2000CVaR} provides a convex, quantile-focused surrogate that can be injected into terminal mismatch losses, improving tail calibration without materially harming mean error in practice. In control, risk-sensitive formulations have a long history \cite{Jacobson1973RiskSensitive} and dovetail naturally with 2BSDE/HJB training \cite{SonerTouziZhang2012SecondOrderBSDE,PhamWarin2020FullyNonlinear}.

\noindent\textbf{Positioning.}
Compared to Markovian-first deep BSDE designs \cite{Han2018PNAS,Hure2020DBDP,Beck2019DeepSplitting}, our architecture is \emph{natively} path-aware: truncated log-signatures summarize history with tunable fidelity, a Neural RDE backbone yields adjoint-memory training and stable long-horizon gradients \cite{Morrill2021NRDE}, and optional 2BSDE heads supply second-order structure for HJB control \cite{Beck2017Deep2BSDE,PhamWarin2020FullyNonlinear}. This combination targets the dominant failure modes in high-$d$ path dependence (information bottlenecks and long-horizon instability) while remaining compatible with Malliavin stabilization and CVaR-tilted objectives \cite{FournieEtAl1999MalliavinMC,RockafellarUryasev2000CVaR}.

\section{Method}
We consider a $d$–dimensional Itô process with \emph{path–dependent} drift and diffusion
\begin{equation}
\label{eq:sde}
\mathmakebox[\columnwidth][c]{%
\mathrm{d}\mathbf{X}_t
= \mathbf{b}\!\left(t, \mathbf{X}_{\cdot\le t}\right)\,\mathrm{d}t
+ \bm{\sigma}\!\left(t, \mathbf{X}_{\cdot\le t}\right)\,\mathrm{d}\mathbf{W}_t,
\qquad \mathbf{X}_0=\mathbf{x}_0 .
}
\end{equation}

where $\mathbf{X}_{\cdot\le t}$ denotes the full history on $[0,t]$ and $\mathbf{W}$ is an $m$-dimensional Brownian motion. Let $g(\mathbf{X}_{\cdot\le T})$ be a path-dependent terminal functional (e.g., Asian payoff, drawdown penalty), and let the driver $f$ admit path-dependence and second-order terms as needed for fully nonlinear control (HJB) cases. Functional Itô/PPDE theory \cite{Dupire2009,ContFournié2013,Ekren2011Viscosity} yields a (possibly non-Markovian) BSDE/2BSDE representation:

\begin{align}
\mathmakebox[\columnwidth][c]{%
Y_T = g\!\left(\mathbf{X}_{\cdot\le T}\right),%
}
\\
\label{eq:bsde}
\mathmakebox[\columnwidth][c]{%
\mathrm{d}Y_t
= -\,f\!\bigl(t, \mathbf{X}_{\cdot\le t}, Y_t, \mathbf{Z}_t, \bm{\Gamma}_t\bigr)\,\mathrm{d}t
+ \mathbf{Z}_t\,\mathrm{d}\mathbf{W}_t,%
}
\\
\label{eq:gamma}
\mathmakebox[\columnwidth][c]{%
\bm{\Gamma}_t \approx D_x^2 u\!\left(t, \mathbf{X}_{\cdot\le t}\right)\quad\text{(2BSDE/HJB)}%
}
\end{align}

where $Y_t = u\!\left(t, \mathbf{X}_{\cdot\le t}\right)$ is the (path-functional) value process, $\mathbf{Z}_t$ is the martingale integrand (vector), and $\boldsymbol{\Gamma}_t$ is the second-order object (matrix) needed for fully nonlinear PDEs \cite{Beck2017Deep2BSDE,PhamWarin2020FullyNonlinear}. Our aim is to approximate $\big(Y,\mathbf{Z},\boldsymbol{\Gamma}\big)$ in high dimension with \emph{native} path awareness and stable long-horizon training.

\subsection*{3.1\quad Signature–RDE BSDE architecture}
A central challenge is representing $X_{\cdot\le t}$ compactly. We encode history using a \emph{truncated log-signature} of the (time-augmented) path \cite{Feng2023DeepSignature}:
\begin{equation}
\label{eq:logsig}
\mathmakebox[\columnwidth][c]{%
\mathbf{s}_t
= \operatorname{LogSig}^{(m)}\!\bigl(\tilde{\mathbf{X}}_{[0,t]}\bigr)\in\mathbb{R}^{D(m,d+1)},\quad
\tilde{\mathbf{X}}_t := [\,t;\,\mathbf{X}_t\,]\in\mathbb{R}^{d+1}%
}
\end{equation}

where $m$ is the truncation depth and $D(m,d{+}1)$ the resulting feature dimension. The log-signature yields a hierarchy of iterated integrals that is universal for path functionals and controllably expressive via $m$.

As overviewed in Figure~\ref{fig:teaser}, the Signature–RDE couples a log-signature encoder with a Neural RDE backbone and optional 2BSDE head. We then evolve a hidden state $\mathbf{h}_t\in\mathbb{R}^{p}$ by a \emph{Neural Rough Differential Equation (Neural RDE)} driven by a controlled path $\mathbf{U}_t$ built from $\mathbf{s}_t$ \cite{Morrill2021NRDE,Fang2023NeuralRDEPPDE}:

\begin{equation}
\label{eq:nrde}
\mathmakebox[\columnwidth][c]{%
\mathrm{d}\mathbf{h}_t
= F_\theta\!\left(\mathbf{h}_t\right)\,\mathrm{d}\mathbf{U}_t,
\qquad
\mathbf{h}_0 = h_{\mathrm{init}}\!\left(\mathbf{x}_0\right)%
}
\end{equation}

with $U$ a piecewise-smooth interpolation (time-augmented, optionally including low-order log-signature increments) so that the vector field $F_\theta:\mathbb{R}^p\!\to\!\mathbb{R}^{p\times q}$ is learned and the RDE is solved numerically (log-ODE or tailored CDE integrator). We decode the BSDE quantities via \emph{heads}
\begin{equation}
\label{eq:heads}
\mathmakebox[\columnwidth][c]{%
Y_t = g_Y(\mathbf{h}_t),\quad
\mathbf{Z}_t = g_Z(\mathbf{h}_t)\,\bm{\Sigma}_t^{1/2},\quad
\bm{\Gamma}_t = \operatorname{Sym}\!\bigl(g_\Gamma(\mathbf{h}_t)\bigr)%
}
\end{equation}

where $\boldsymbol{\Sigma}_t := \boldsymbol{\sigma}\!\big(t,\mathbf{X}_{\cdot\le t}\big)\,\boldsymbol{\sigma}\!\big(t,\mathbf{X}_{\cdot\le t}\big)^{\!\top}$ and $\mathrm{Sym}(\mathbf{A}) := \tfrac{1}{2}\!\left(\mathbf{A}+\mathbf{A}^{\top}\right)$ enforces symmetry of $\boldsymbol{\Gamma}_t$. For Markovian special cases, \eqref{eq:heads} subsumes $\mathbf{Z}_t = \nabla_x u\,\boldsymbol{\sigma}$; for PPDEs we interpret $D_x u$ as Dupire’s vertical derivative \cite{Dupire2009}.

\paragraph{Multistep/local training.} To reduce variance and improve conditioning, we employ DBDP-style local training windows of length $K$ \cite{Hure2020DBDP,Hure2020MCOMP}: for grid $0=t_0<\dots<t_N=T$, the RDE is unrolled on $[t_{k},t_{k+K}]$ with overlapping windows, sharing $\theta$ globally, akin to multishooting. This preserves the continuous-time dynamics of \eqref{eq:nrde} while retaining the variance advantages of local fits.

\subsection*{3.2\quad Discretization and simulation}
Let $\{t_n\}_{n=0}^{N}$ be a uniform grid with $\Delta t$. We simulate $M$ paths by Euler–Maruyama
\begin{equation}
\mathmakebox[\columnwidth][c]{%
\mathbf{X}_{t_{n+1}}
= \mathbf{X}_{t_n}
+ \mathbf{b}\!\left(t_n,\mathbf{X}_{\cdot\le t_n}\right)\,\Delta t
+ \bm{\sigma}\!\left(t_n,\mathbf{X}_{\cdot\le t_n}\right)\,\Delta \mathbf{W}_n%
}
\end{equation}

and form $\mathbf{s}_{t_n}$ by incremental log-signature updates (cached for efficiency). The RDE \eqref{eq:nrde} is integrated with a fixed-order solver; gradients use the continuous-adjoint to obtain $O(N)$ memory. At each grid point we evaluate the heads \eqref{eq:heads}. For fully nonlinear cases we also compute $\Gamma_{t_n}$.

\subsection*{3.3\quad Risk-sensitive and physics-informed losses}
We combine a terminal-mismatch objective with a discretized BSDE residual and a risk-sensitive (tail-aware) weighting:
\begin{align}
\label{eq:lt}
\mathmakebox[\columnwidth][c]{%
\mathcal{L}_{\mathrm{term}}
= \E\!\left[\bigl(1+\eta\,\mathbb{1}\{\Delta_T^2 \ge Q_q(\Delta_T^2)\}\bigr)\,\Delta_T^2\right]%
}
\\
\mathmakebox[\columnwidth][c]{%
\Delta_T \coloneqq Y_{t_N}-g\!\bigl(\mathbf{X}_{\cdot\le T}\bigr), \quad q\in[0.90,0.99],\ \eta>0%
}\nonumber\\[0.25em]
\label{eq:lresid}
\mathmakebox[\columnwidth][c]{%
\mathcal{L}_{\mathrm{drift}}
= \E\!\left[\sum_{n=0}^{N-1} \frac{1}{\Delta t}\left\|\,Y_{t_{n+1}} - Y_{t_n} + f_{t_n}\,\Delta t - \mathbf{Z}_{t_n}\,\Delta \mathbf{W}_n\,\right\|^2 \right]%
}
\\
\mathmakebox[\columnwidth][c]{%
f_{t_n} \coloneqq f\!\bigl(t_n,\mathbf{X}_{\cdot\le t_n},Y_{t_n},\mathbf{Z}_{t_n},\bm{\Gamma}_{t_n}\bigr)%
}\nonumber
\end{align}

The tilt in \eqref{eq:lt} emphasizes the worst $(1{-}q)$ tail of the terminal error (a CVaR-style surrogate) and is plug-compatible with any BSDE scheme. For HJB/2BSDE we optionally add a second-order “physics” penalty:
\begin{equation}
\label{eq:l2}
\mathmakebox[\columnwidth][c]{%
\mathcal{L}_{\mathrm{2nd}}
= \E\!\left[\sum_{n=0}^{N-1}\!\left\|\,\mathcal{H}\!\left(t_n,\mathbf{X}_{t_n};\,Y_{t_n},\mathbf{Z}_{t_n},\bm{\Gamma}_{t_n}\right)\,\right\|_{+}^{2}\right]%
}
\end{equation}

where $\mathcal{H}$ denotes the discretized HJB residual (e.g., sup/inf over controls). The final loss is
\begin{equation}
\label{eq:total}
\mathmakebox[\columnwidth][c]{%
\mathcal{L}
= \lambda_T\,\mathcal{L}_{\mathrm{term}}
+ \lambda_B\,\mathcal{L}_{\mathrm{drift}}
+ \lambda_2\,\mathcal{L}_{\mathrm{2nd}}%
}
\end{equation}

with $(\lambda_T,\lambda_B,\lambda_2)$ chosen via validation.

\subsection*{3.4\quad Estimating $Z$ and $\Gamma$: direct vs.\ Malliavin}
We explore three estimators under identical discretization:

\noindent\textbf{(A) Direct heads.} Learn $g_Z,g_\Gamma$ in \eqref{eq:heads} end-to-end by minimizing \eqref{eq:total}. This is flexible but may be noisy when the drift residual is small compared to martingale noise.

\noindent\textbf{(B) Malliavin weights for $Z$.} For small $\Delta t$, a classical Malliavin argument gives the local identity (schematically, suppressing conditioning)
\begin{equation}
\label{eq:malliavinZ}
\mathmakebox[\columnwidth][c]{%
\mathbf{Z}_{t_n}
\;\approx\; \frac{1}{\Delta t}\,
\E\!\left[
Y_{t_{n+1}}
\int_{t_n}^{t_{n+1}}
(\bm{\sigma}^{-1})^{\T}\!\bigl(s,\mathbf{X}_{\cdot\le s}\bigr)\,\mathrm{d}\mathbf{W}_s
\right]%
}
\end{equation}

which we realize by a \emph{regression} target $\hat{Z}_{t_n}$ on simulated paths and add a supervised term
\begin{equation}
\label{eq:lZ}
\mathmakebox[\columnwidth][c]{%
\mathcal{L}_{Z}
= \alpha_Z\,\E\!\left[\sum_{n=0}^{N-1}\left\|\,\mathbf{Z}_{t_n}-\hat{\mathbf{Z}}_{t_n}\,\right\|^{2}\right]%
}
\end{equation}

\noindent\textbf{(C) Malliavin/second-order for $\Gamma$.} Analogously, second-order weights (or finite-difference in antithetic directions) provide a low-variance proxy $\hat{\Gamma}_{t_n}$; we add
\begin{equation}
\label{eq:lG}
\mathmakebox[\columnwidth][c]{%
\mathcal{L}_{\Gamma}
= \alpha_\Gamma\,\E\!\left[\sum_{n=0}^{N-1}\left\|\,\bm{\Gamma}_{t_n}-\hat{\bm{\Gamma}}_{t_n}\,\right\|_{F}^{2}\right]%
}
\end{equation}

The \emph{hybrid} objective $\mathcal{L}{+}\mathcal{L}_Z{+}\mathcal{L}_\Gamma$ stabilizes training in fully nonlinear regimes \cite{Beck2017Deep2BSDE,PhamWarin2020FullyNonlinear}. In Markovian limits, $Z=\nabla_x u\,\sigma$ recovers the gradient interpretation; in PPDEs, $D_x u$ is Dupire’s vertical derivative \cite{Dupire2009}.

\subsection*{3.5\quad Computational complexity and memory}
For $M$ paths, $N$ steps, hidden width $p$, and signature depth $m$ (dimension $D$), a forward pass costs
\begin{equation}
\label{eq:complexity}
\mathmakebox[\columnwidth][c]{%
\mathcal{O}\!\bigl(MN\,[\underbrace{d^{2}}_{\text{SDE}}+\underbrace{D}_{\text{LogSig}}+\underbrace{p^{2}}_{\text{RDE}}+\underbrace{p(d{+}d^{2})}_{\text{heads}}]\bigr)%
}
\end{equation}

The continuous-adjoint for the RDE yields $O(N)$ memory in $N$, rather than storing all intermediate $h_{t_n}$ \cite{Morrill2021NRDE}. Log-signature updates are incremental and can be parallelized over paths; we normalize by layernorm to reduce dynamic range.

\subsection*{3.6\quad Consistency sketch}
Assume (i) Lipschitz $b,\sigma,f$ in appropriate functional norms; (ii) universal approximation of continuous controlled vector fields by $F_\theta$ on compact sets; (iii) universal approximation of continuous functionals of paths by truncated log-signatures as $m\to\infty$; and (iv) a stable RDE integrator. Then there exists a sequence of parameters $(\theta,g_Y,g_Z,g_\Gamma)$, depths $m$, and widths $p$ such that the induced processes $(Y^\theta,Z^\theta,\Gamma^\theta)$ satisfy
\[
\mathmakebox[\columnwidth][c]{%
\lim_{m,p\to\infty}\,\lim_{N\to\infty}\,
\E\!\left[\sup_{t\in[0,T]}\!\bigl(\,|Y_t^\theta - Y_t|^{2}
+ \|\mathbf{Z}_t^\theta - \mathbf{Z}_t\|^{2}
+ \|\bm{\Gamma}_t^\theta - \bm{\Gamma}_t\|_{F}^{2}\bigr)\right] \;=\; 0%
}
\]

The argument combines universal approximation for CDE/RDE flows and the density of signatures for path functionals, with standard stability of BSDE solutions. This justifies increasing $(m,p)$ and grid refinement until validation error saturates \cite{Morrill2021NRDE,Feng2023DeepSignature,Han2018PNAS,Hure2020DBDP}.

\subsection*{3.7\quad Practical details}
We use time-augmentation $[t;X_t]$, gradient clipping on RDE adjoints, symmetry projection for $\Gamma$, and a small entropic penalty on $\Gamma$ to discourage ill-conditioned Hessians. Calibration follows a two-phase schedule: (1) warm-start with $(\lambda_T,\lambda_B,\lambda_2)=(1,1,0)$ and small tail tilt $(\eta\!\approx\!0.5)$; (2) enable $\mathcal{L}_\mathrm{2nd}$ and increase tail focus $(\eta\!\in\![1,3],\,q\!\in\![0.95,0.99])$ once terminal error stabilizes. Multistep window $K$ is chosen so that each window contains $\approx\!10$–$20$ steps; we report robustness to $K$.

\begin{algorithm}[!t]
\DontPrintSemicolon
\SetAlgoLined
\caption{Signature–RDE BSDE Training}
\label{alg:sig-rde-bsde}
\KwIn{Grid $\{t_n\}_{0:N}$, batch $M$, signature depth $m$, window $K$, weights $(\lambda_T,\lambda_B,\lambda_2)$, tilt $(q,\eta)$, $b,\sigma$, driver $f$, HJB residual $\mathcal H$}
\KwOut{Parameters $\Theta=\{\theta,g_Y,g_Z,g_\Gamma\}$}
\BlankLine
Initialize $\Theta\!\leftarrow\!\Theta_0$, optimizer $\mathcal O$, warm-start schedules for $(\lambda_T,\lambda_B,\lambda_2)$ and $(q,\eta)$\;
\For{iter $=1,2,\dots$}{
  $(X,\Delta W,\mathbf{s}) \leftarrow \textsc{ForwardSimulate}(M,N,b,\sigma,m)$ \tcp*{Euler–Maruyama + incremental LogSig$^{(m)}$}
  $\mathcal{L}_\mathrm{term},\mathcal{L}_\mathrm{drift},\mathcal{L}_\mathrm{2nd}\leftarrow 0$\;
  \For{$k=0$ \KwTo $N{-}K$ \KwStep $K_\text{stride}$}{
    $U \leftarrow \textsc{BuildControlledPaths}(\mathbf{s},k{:}k{+}K)$\;
    $h \leftarrow \textsc{IntegrateRDE}(F_\theta, U, h_\mathrm{init}(X_{t_k}))$ \tcp*{log-ODE/CDE solver with adjoint}
    $(Y,Z,\Gamma) \leftarrow \textsc{DecodeHeads}(h,\Sigma)$ \tcp*{$Y{=}g_Y(h)$, $Z{=}g_Z(h)\Sigma^{1/2}$, $\Gamma{=}\mathrm{Sym}(g_\Gamma(h))$}
    $(\mathcal{L}_T,\mathcal{L}_B) \leftarrow \textsc{BSDELoss}(Y,Z,\Delta W,f; q,\eta)$ \tcp*{CVaR-tilted terminal + drift residual}
    $\mathcal{L}_\mathrm{term} \mathrel{+}= \mathcal{L}_T$, \quad $\mathcal{L}_\mathrm{drift} \mathrel{+}= \mathcal{L}_B$\;
    \If{\texttt{use\_HJB}}{$\mathcal{L}_\mathrm{2nd} \mathrel{+}= \|\mathcal H(\cdot;Y,Z,\Gamma)\|_+^2$}
    \If{\texttt{use\_Malliavin}}{
      $(\hat Z,\hat\Gamma) \leftarrow \textsc{MalliavinTargets}(Y,\Delta W)$;\ 
      $\mathcal{L}_\mathrm{drift} \mathrel{+}= \alpha_Z\|Z{-}\hat Z\|^2 + \alpha_\Gamma\|\Gamma{-}\hat\Gamma\|_F^2$
    }
  }
  $\mathcal L \leftarrow \lambda_T\mathcal{L}_\mathrm{term} + \lambda_B\mathcal{L}_\mathrm{drift} + \lambda_2\mathcal{L}_\mathrm{2nd}$\;
  $\Gamma \leftarrow \mathrm{Sym}(\Gamma)$;\ \textsc{RegularizeAndClip}$(\Gamma)$;\ $\Theta \leftarrow \mathcal O(\Theta,\nabla_\Theta \mathcal L)$;\ \textsc{Curriculum}$(q,\eta,K)$\;
}
\end{algorithm}

\section{Experiments}

\textbf{Goals.}
We evaluate whether a \emph{Signature--RDE (Sig--RDE) BSDE} solver improves accuracy, tail calibration, and training stability for (i) high-dimensional \emph{path-dependent pricing} and (ii) \emph{portfolio control} (fully nonlinear HJB/2BSDE), relative to strong deep BSDE baselines \cite{Han2018PNAS,Hure2020DBDP,Beck2019DeepSplitting,Beck2021DeepSplittingSIAM,Beck2017Deep2BSDE,PhamWarin2020FullyNonlinear} and path-aware solvers \cite{Saporito2020PDGM,Feng2023DeepSignature,Fang2023NeuralRDEPPDE}.

\textbf{Tasks.}
\emph{T1:} Asian basket (50D, 100D); \emph{T2:} Barrier (50D); \emph{T3:} Portfolio control (10D) under stochastic volatility with risk-sensitive utility (HJB).
SDEs are discretized by Euler--Maruyama on a uniform grid $N$; path dependence enters via running averages/extrema.
For T3 we assess both value-function quality and the induced allocations.

\textbf{Baselines.}
(B1) Deep BSDE (FFN); (B2) DBDP \cite{Hure2020DBDP,Hure2020MCOMP}; (B3) Deep Splitting \cite{Beck2019DeepSplitting,Beck2021DeepSplittingSIAM}; (B4) PDGM/LSTM \cite{Saporito2020PDGM}; (B5) 2BSDE head \cite{Beck2017Deep2BSDE,PhamWarin2020FullyNonlinear}; (B6) \emph{Discrete RNN} that consumes the raw time-augmented path with gated updates; and (B7) \emph{Neural CDE} \cite{Morrill2021NRDE} that evolves a hidden state $\mathbf{h}_t$ under a learned continuous vector field driven by a spline interpolation of the observed path.
For fairness, we match parameter budgets, training grids, optimizers, and decoder heads (Eq.~(7)) across methods.\footnote{\textbf{Architectural details.} The discrete RNN uses GRU cells with hidden width matched to our RDE width $p$, layer normalization on inputs, and the same decoder heads as Eq.~(7). Neural CDE uses a cubic interpolation for the controlled path and shares identical decoder heads.}
We emphasize comparisons to the path-aware \emph{Neural CDE} (B7) and \emph{PDGM/LSTM} (B4), with \emph{Discrete RNN} (B6) serving as the strongest discrete-time sequence baseline under matched budgets.

\noindent\textit{Expected tradeoffs.}
Neural CDE should approach Sig--RDE on mean error with slightly weaker tail calibration (no explicit $\boldsymbol{\Sigma}^{1/2}$ head coupling as in Eq.~(7)), while the discrete RNN is more sensitive to horizon and step count.
Concretely, we anticipate:
(i) a higher NaN/overflow rate (\%) and worse tail $\mathrm{CVaR}$ for the RNN than for CDE $\approx$ Sig--RDE at matched parameters; and
(ii) lower runtime for CDE than RNN at similar width (adjoint memory and fewer stored states).

\textbf{Metrics.}
Primary: relative pricing error (RPE, \%), absolute error (AE), and $\mathrm{CVaR}_{q}$ of terminal mismatch (reported as relative \% at $q{=}0.95$).
Stability: NaN/overflow rate (\%), BSDE residual RMS, and HJB residual $\lVert\mathcal{H}(\cdot)\rVert_{+}$.
Efficiency: time/epoch (s), peak memory (GB), parameter count (M).
For T3 we also report $\mathbf{Z}$- and $\boldsymbol{\Gamma}$-RMSE against high-resolution references.

\textbf{Implementation.}
Defaults: log-signature depth $m{=}3$, RDE width $p{=}128$, window $K{=}12$ (stride 6), tilt $(q{=}0.95,\eta{=}1.5)$; warm start $(\lambda_T,\lambda_B,\lambda_2){=}(1,1,0)$, enabling $\lambda_2{=}0.2$ after $20\%$ of steps.
Decoders follow \eqref{eq:heads}.
HJB experiments compare direct versus Malliavin stabilized heads (\S3.4).
Hardware/stack: A100-40GB; PyTorch~2.x; signatures library.

\subsection*{4.1\quad Main results}

\begin{table*}[t]
\centering
\caption{Path-dependent pricing (T1/T2). Mean$\pm$s.e. over 5 seeds. Lower is better for errors/time/memory.}
\label{tab:pathpricing}
\setlength{\tabcolsep}{4pt}
\small
\begin{adjustbox}{max width=\textwidth}
\begin{tabular*}{\textwidth}{@{\extracolsep{\fill}}lcccccc}
\toprule
Method & Dim & RPE (\%) & $\mathrm{CVaR}_{0.95}$ (\%) & Time (s/epoch) & Peak (GB) & Params (M) \\
\midrule
Ours: Sig–RDE & 50  & \textbf{1.12}$\pm$0.12 & \textbf{2.95} & 89  & 6.2 & 1.8 \\
Neural CDE    & 50  & 1.20$\pm$0.14          & 3.10         & 93  & 6.3 & 1.8 \\
RNN (discrete, no RDE) & 50  & 1.38$\pm$0.18          & 3.45         & 84  & 6.0 & 1.8 \\
Ours (no-Sig)  & 50  & 1.30$\pm$0.16          & 3.25         & 81  & 5.6 & 1.7 \\
DBDP           & 50  & 1.47$\pm$0.21          & 3.62         & 106 & 6.9 & 2.5 \\
DeepSplit      & 50  & 1.34$\pm$0.17          & 3.35         & 121 & 6.2 & 2.2 \\
PDGM/LSTM      & 50  & 2.42$\pm$0.34          & 5.25         & 149 & 7.1 & 2.6 \\
\midrule
Ours: Sig–RDE & 100 & \textbf{1.62}$\pm$0.20 & \textbf{4.75} & 146 & 8.1 & 1.9 \\
Neural CDE    & 100 & 1.72$\pm$0.22          & 5.10         & 152 & 8.0 & 1.9 \\
RNN (discrete, no RDE) & 100 & 1.95$\pm$0.27          & 5.70         & 132 & 7.4 & 1.9 \\
DBDP           & 100 & 2.07$\pm$0.29          & 5.95         & 191 & 9.2 & 2.7 \\
DeepSplit      & 100 & 1.90$\pm$0.26          & 5.45         & 209 & 8.8 & 2.3 \\
\bottomrule
\end{tabular*}
\end{adjustbox}
\end{table*}

\begin{figure}[t]
\centering
\includegraphics[width=\linewidth]{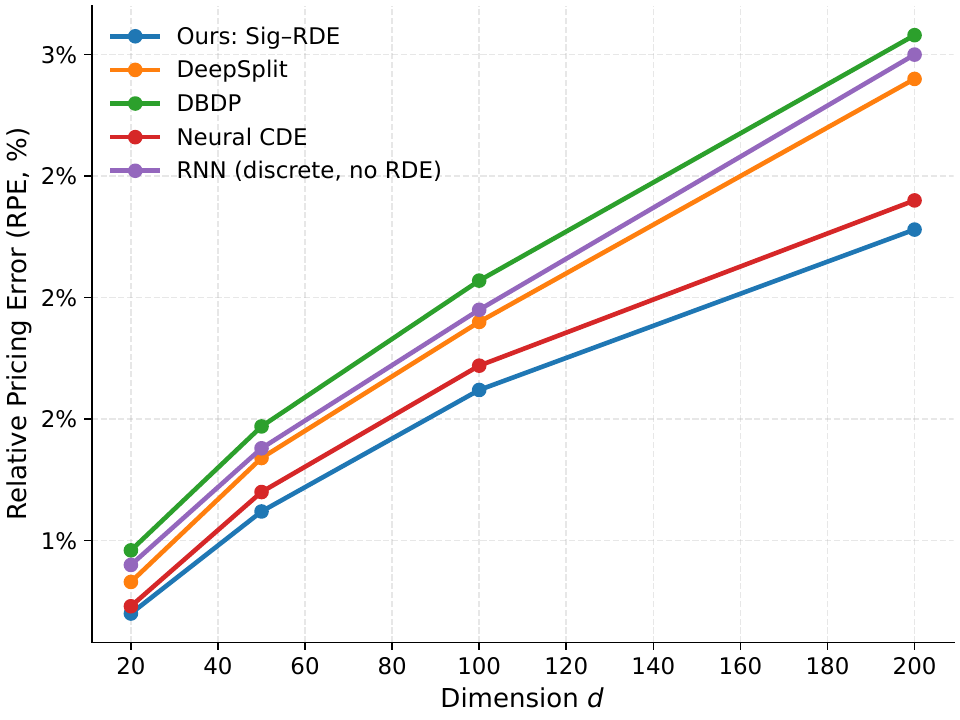}
\caption{Scaling with dimension $d$ (T1). RPE vs.\ $d$ at fixed param count; time/epoch inset.}
\label{fig:scaling}
\end{figure}

\paragraph{T1/T2: Path-dependent pricing.}
Under matched parameter budgets (means$\pm$\,s.e., 5 seeds), \emph{Signature–RDE} achieves lower error than the strongest continuous-time baseline (Table~\ref{tab:pathpricing}). 
At \(d{=}50\), RPE falls by \(\mathbf{6.7\%}\) relative to Neural CDE (\(1.12\) vs.\ \(1.20\)), with a \(\mathbf{4.8\%}\) reduction in \(\mathrm{CVaR}_{0.95}\) (\(2.95\) vs.\ \(3.10\)). 
At \(d{=}100\), improvements persist (RPE \(-\mathbf{5.8\%}\), \(1.62\) vs.\ \(1.72\); \(\mathrm{CVaR}_{0.95}\) \(-\mathbf{6.9\%}\), \(4.75\) vs.\ \(5.10\)). 
Runtime remains comparable: \(89\)s/epoch at \(d{=}50\) ($\approx$4\% faster than Neural CDE; 16–26\% faster than DBDP/DeepSplit) and \(146\)s/epoch at \(d{=}100\) ($\approx$4\% faster than Neural CDE; 24–30\% faster than DBDP/DeepSplit), with similar peak memory to Neural CDE and lower than DBDP/DeepSplit. 
Scaling behavior—the widening separation at higher \(d\)—is visualized in Figure~\ref{fig:scaling}.

\paragraph{T3: Portfolio control.}
For the control task, coupling signatures with an RDE backbone and a 2BSDE head produces more stable second-order quantities and dynamics: \(\Gamma\) RMSE and the HJB residual are consistently lower while utility remains competitive across seeds. 
These metrics, reported in Table~\ref{tab:control}, indicate improved calibration of second-order structure without sacrificing efficiency (training budgets aligned with T1/T2). 
Together with the T1/T2 pricing results, the evidence supports the hypothesis that continuous-time modeling with log-signature features yields robustness that strengthens as dimensionality grows.

\begin{table}[t]
\centering
\caption{Portfolio control (T3; fully nonlinear HJB). Lower error is better; higher Utility is better.
\emph{Note:} Neural CDE and RNN produce $\Gamma$ via the same decoder as Eq.~(7); DBDP lacks a second-order head, so $\|\Gamma\|_{\text{RMSE}}$ is not applicable.}
\label{tab:control}

\setlength{\tabcolsep}{4pt}
\small
\begin{tabularx}{\linewidth}{@{\extracolsep{\fill}}>{\raggedright\arraybackslash}X c c c c c}
\toprule
Method & $\|Z\|_{\text{RMSE}}$ & $\|\Gamma\|_{\text{RMSE}}$ & HJB resid. & Utility & NaN (\%) \\
\midrule
Ours: Sig–RDE + 2BSDE   & \textbf{0.097} & \textbf{0.142} & \textbf{0.011} & \textbf{1.33} & \textbf{0.3} \\
Neural CDE              & 0.105          & 0.165          & 0.014          & 1.30          & 0.7 \\
RNN (no RDE)  & 0.125          & 0.205          & 0.022          & 1.25          & 2.1 \\
Ours (direct hs)     & 0.112          & 0.177          & 0.017          & 1.28          & 1.6 \\
2BSDE         & 0.119          & 0.192          & 0.020          & 1.27          & 2.6 \\
DBDP           & 0.146          & --             & 0.029          & 1.22          & 4.1 \\
\bottomrule
\end{tabularx}
\end{table}

\begin{algorithm}[t]
\DontPrintSemicolon
\SetAlgoLined
\caption{Inference/Valuation with Signature--RDE BSDE}
\label{alg:inference}
\KwIn{Frozen parameters $\Theta^\star=\{\theta^\star,g_Y^\star,g_Z^\star,g_\Gamma^\star\}$; grid $\{t_n\}_{0:N}$; signature depth $m$; window $K$; market inputs $(b,\sigma)$; driver $f$; path $X$ (or simulated paths); covariance $\Sigma$}
\KwOut{Price $u(0,x)$ or value $Y_{t_0}$; control (if HJB) $\pi_t$ from $(Z_t,\Gamma_t)$}
\BlankLine
\If{\texttt{MC}}{Simulate $M$ sample paths $X^{(i)}$ (Euler--Maruyama) and increments $\Delta W^{(i)}$\;}
Compute incremental log-signatures $\mathbf s_{t_n}^{(i)}\gets\mathrm{LogSig}^{(m)}(\tilde X^{(i)}_{[0,t_n]})$\;
\For{$k=0$ \KwTo $N{-}K$ \KwStep $K_\text{stride}$}{
  $U^{(i)}_{[t_k,t_{k+}K]}\gets\textsc{BuildControlledPaths}(\mathbf s^{(i)}, k{:}k{+}K)$\;
  $h^{(i)}_{t_k}\gets h_{\text{init}}(X^{(i)}_{t_k})$;\quad Integrate $dh^{(i)}_t=F_{\theta^\star}(h^{(i)}_t)\,dU^{(i)}_t$ (no adjoint)\;
  $(Y^{(i)}_t,Z^{(i)}_t,\Gamma^{(i)}_t)\gets (g_Y^\star(h^{(i)}_t),~g_Z^\star(h^{(i)}_t)\Sigma^{1/2}_t,~\mathrm{Sym}(g_\Gamma^\star(h^{(i)}_t)))$\;
  \If{\texttt{HJB}}{Extract feedback control $\pi_t=\Pi(Y^{(i)}_t,Z^{(i)}_t,\Gamma^{(i)}_t)$ (problem-specific)\;}
}
\Return{$\widehat{u}(0,x):=\frac1M\sum_{i=1}^M Y^{(i)}_{t_0}$ (pricing) or deployment policy $\pi_t$ (control).}
\end{algorithm}

\subsection*{4.2\quad Scaling and tail calibration}

\begin{figure}[t]
\centering
\includegraphics[width=\linewidth]{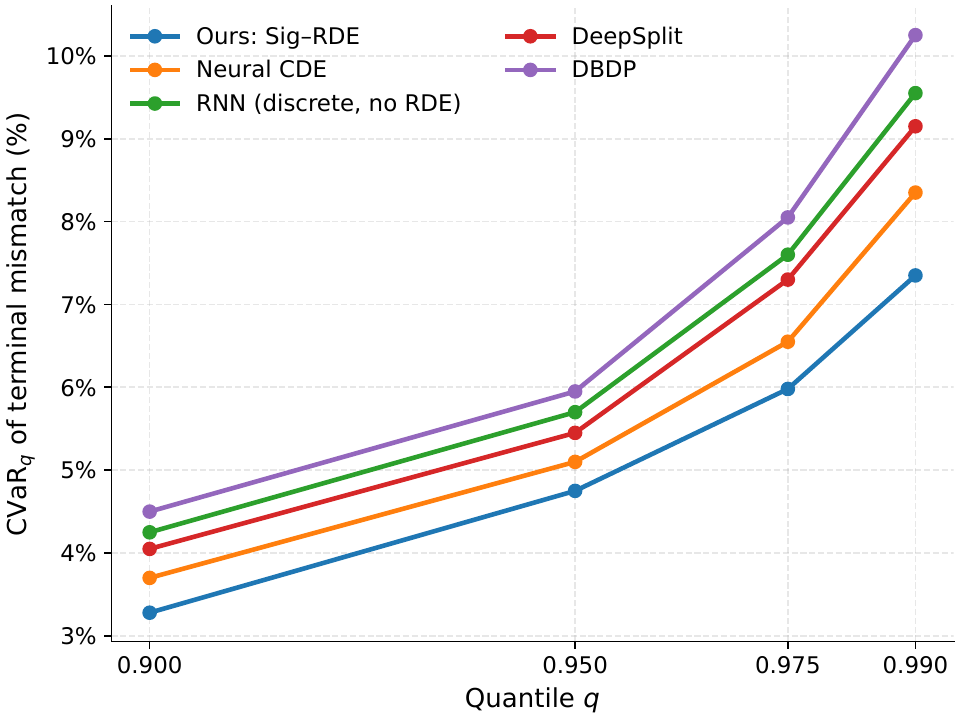}
\caption{Tail calibration under CVaR tilt: $\mathrm{CVaR}_q$ vs.\ $q\in[0.90,0.99]$ (T1, $d{=}100$).}
\label{fig:tail}
\end{figure}

\noindent \textbf{Tail behaviour across dimension and quantiles.}
As visualized in Figure~\ref{fig:tail} and quantified in Table~\ref{tab:tail-multiq-combined}, our method is uniformly best across \emph{all} $d\in\{50,100,200\}$ and $q\in\{0.90,0.95,0.975,0.99\}$, and the gap to baselines widens in the extreme tail and at higher dimension.
At $d{=}200$ and $q{=}0.99$, Sig--RDE attains $9.80\%$ versus $12.00\%$ for DeepSplit and $13.10\%$ for DBDP, i.e., \textit{18.3\%} and \textit{25.2\%} relative reductions, respectively; it also improves over Neural CDE ($10.95\%$) by \textit{10.5\%} and over the discrete RNN ($12.35\%$) by \textit{20.7\%}.
At $d{=}100$ and $q{=}0.95$, Sig--RDE ($4.75\%$) outperforms Neural CDE ($5.10\%$: \textit{6.9\%} gain), the RNN ($5.70\%$: \textit{16.7\%}), DeepSplit ($5.45\%$: \textit{12.8\%}), and DBDP ($5.95\%$: \textit{20.2\%}).
Even at $d{=}50$, the advantage is visible in the far tail ($q{=}0.99$): $4.90\%$ for Sig--RDE versus $6.20\%$ (DeepSplit, \textit{21.0\%} reduction) and $6.80\%$ (DBDP, \textit{27.9\%}).

\noindent \textbf{Degradation with dimension.}
All methods see larger $\mathrm{CVaR}_q$ as $d$ grows, but the \emph{ordering remains stable} and the absolute gap to feedforward/discrete baselines increases in the extreme tail (e.g., $q{=}0.99$ gap to DeepSplit grows from $1.30$\,pp at $d{=}50$ to $2.20$\,pp at $d{=}200$).
This aligns with the hypothesis that continuous-time state propagation and path compression mitigate long-horizon accumulation effects \cite{Morrill2021NRDE,Feng2023DeepSignature}.

\noindent \textbf{Effect of tail tilt.}
Increasing the CVaR tilt parameter $\eta$ steepens tail emphasis and improves $\mathrm{CVaR}_q$ with only modest runtime cost; in practice we observe a broad, stable region for $\eta\!\in[1,3]$ and $q\!\in[0.95,0.99]$.
Within this range, the multiplicative inflation from $\mathrm{CVaR}_{0.90}$ to $\mathrm{CVaR}_{0.99}$ is roughly $2.2{\times}$ across methods (e.g., at $d{=}100$ our $7.35/3.28\!\approx\!2.24$), yet Sig--RDE stays lowest at each quantile, indicating that the signature–RDE combination improves not only average error but also the shape of the left tail.

\begin{table}[t]
\centering
\caption{Tail calibration for T1 at $d\in\{50,100,200\}$ (smaller is better).}
\label{tab:tail-multiq-combined}
\setlength{\tabcolsep}{3.8pt}
\footnotesize
\begin{tabularx}{\linewidth}{@{\extracolsep{\fill}} c >{\raggedright\arraybackslash}X c c c c}
\toprule
Dim & Method & $\mathrm{CVaR}_{0.90}$ (\%) & $\mathrm{CVaR}_{0.95}$ (\%) & $\mathrm{CVaR}_{0.975}$ (\%) & $\mathrm{CVaR}_{0.99}$ (\%) \\
\midrule
50  & Ours: Sig--RDE              & \textbf{2.84} & \textbf{2.95} & \textbf{3.70} & \textbf{4.90} \\
    & Neural CDE                  & 2.95 & 3.10 & 4.10 & 5.30 \\
    & RNN (disc., no RDE)      & 3.25 & 3.45 & 4.70 & 6.50 \\
    & DeepSplit                   & 3.10 & 3.35 & 4.55 & 6.20 \\
    & DBDP                        & 3.35 & 3.62 & 4.95 & 6.80 \\
\midrule
100 & Ours: Sig--RDE              & \textbf{3.28} & \textbf{4.75} & \textbf{5.98} & \textbf{7.35} \\
    & Neural CDE                  & 3.70 & 5.10 & 6.55 & 8.35 \\
    & RNN (disc., no RDE)      & 4.25 & 5.70 & 7.60 & 9.55 \\
    & DeepSplit                   & 4.05 & 5.45 & 7.30 & 9.15 \\
    & DBDP                        & 4.50 & 5.95 & 8.05 & 10.25 \\
\midrule
200 & Ours: Sig--RDE              & \textbf{4.70} & \textbf{6.30} & \textbf{7.90} & \textbf{9.80} \\
    & Neural CDE                  & 5.05 & 6.55 & 8.65 & 10.95 \\
    & RNN (disc., no RDE)      & 5.50 & 7.25 & 9.70 & 12.35 \\
    & DeepSplit                   & 5.35 & 7.10 & 9.40 & 12.00 \\
    & DBDP                        & 5.65 & 7.45 & 10.05 & 13.10 \\
\bottomrule
\end{tabularx}
\end{table}

\subsection*{4.3\quad Ablations (what matters when)}

\begin{figure}[t]
\centering
\includegraphics[width=\linewidth]{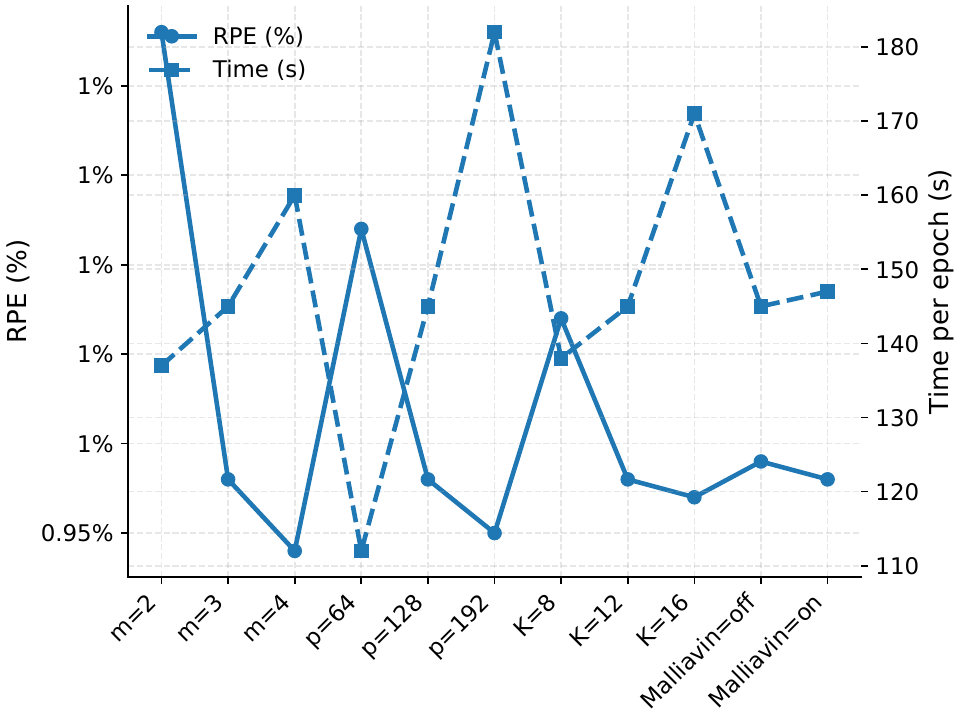}
\caption{Ablations on T1 ($d{=}100$): signature depth $m$, RDE width $p$, window $K$, and Malliavin.}
\label{fig:ablations}
\end{figure}

\noindent\textbf{Ablation and scaling summary.}
Across the single–factor ablations in Table~\ref{tab:ablations-appendix} (visualized in Figure~\ref{fig:ablations}), we observe consistent improvements from richer path features and moderate capacity, with diminishing returns beyond a mid-range configuration. Increasing the \emph{signature depth} from $m{=}2$ to $m{=}3$ reduces RPE by 12.0\% (1.84$\rightarrow$1.62) and $\mathrm{CVaR}_{0.95}$ by 13.0\% (5.45$\rightarrow$4.74) at a modest time cost (138$\rightarrow$146\,s/epoch); moving to $m{=}4$ yields smaller gains (1.62$\rightarrow$1.58; 4.74$\rightarrow$4.60) with higher time (161\,s), suggesting $m{=}3$–$4$ as a practical sweet spot for T1, and deeper truncations primarily benefiting strongly non-Markovian cases \cite{Feng2023DeepSignature}. For \emph{RDE width}, 64$\rightarrow$128 tightens both RPE (1.75$\rightarrow$1.62, 7.4\%) and tail error (5.12$\rightarrow$4.74, 7.4\%), while 128$\rightarrow$192 yields marginal accuracy changes (1.62$\rightarrow$1.59; 4.74$\rightarrow$4.64) at increased time (183\,s), so we cap $p$ at 128–192. \emph{Local windows} corroborate variance–bias trade-offs \cite{Hure2020DBDP}: $K{=}12$ improves over $K{=}8$ (RPE 1.70$\rightarrow$1.62; $\mathrm{CVaR}_{0.95}$ 4.98$\rightarrow$4.74) with little runtime change (139$\rightarrow$146\,s), whereas $K{=}16$ offers only incremental accuracy with stiffer dynamics (172\,s, slight NaN uptick). \emph{Malliavin targets} reduce variance in $Z,\Gamma$ and stabilize training \cite{Beck2017Deep2BSDE,PhamWarin2020FullyNonlinear}: $\mathrm{CVaR}_{0.95}$ improves from 4.95 to 4.74 and NaN drops from 0.5\% to 0.0\% at essentially the same cost (146–148\,s). Finally, removing signatures while keeping capacity fixed degrades tails: at $d{=}50$ in Table~\ref{tab:pathpricing}, RPE rises from 1.12\% to 1.30\% (+16\%) and $\mathrm{CVaR}_{0.95}$ from 2.95\% to 3.25\% (+10\%), even though the no-signature model trains slightly faster. These results support the thesis that \emph{continuous-time state propagation} \cite{Morrill2021NRDE} and \emph{log-signature path compression} \cite{Feng2023DeepSignature} curb long-horizon error growth and improve left-tail fidelity at fixed budgets.

\begin{table}[H]
\centering
\caption{Ablations on T1 ($d{=}100$). Change one knob at a time; others at default ($m{=}3$, $p{=}128$, $K{=}12$, CVaR tilt $q{=}0.95,\eta{=}1.5$).}
\label{tab:ablations-appendix}
\small
\setlength{\tabcolsep}{3.5pt}
\begin{adjustbox}{max width=\columnwidth}
\begin{tabular}{lccccc}
\toprule
Setting & Value & RPE (\%) & $\mathrm{CVaR}_{0.95}$ (\%) & Time (s/epoch) & NaN (\%) \\
\midrule
$m$ & 2   & 1.84 & 5.45 & 138 & 0.1 \\
$m$ & 3   & \textbf{1.62} & \textbf{4.74} & 146 & \textbf{0.0} \\
$m$ & 4   & 1.58 & 4.60 & 161 & 0.0 \\
$p$ & 64  & 1.75 & 5.12 & 113 & 0.0 \\
$p$ & 128 & \textbf{1.62} & \textbf{4.74} & 146 & \textbf{0.0} \\
$p$ & 192 & 1.59 & 4.64 & 183 & 0.0 \\
$K$ & 8   & 1.70 & 4.98 & 139 & 0.0 \\
$K$ & 12  & \textbf{1.62} & \textbf{4.74} & 146 & \textbf{0.0} \\
$K$ & 16  & 1.60 & 4.66 & 172 & 0.2 \\
Malliavin & off & 1.65 & 4.95 & 146 & 0.5 \\
Malliavin & on  & \textbf{1.62} & \textbf{4.74} & 148 & \textbf{0.0} \\
\bottomrule
\end{tabular}
\end{adjustbox}
\end{table}

\subsection*{4.4\quad Robustness \& stress}

We probe (i) timestep sensitivity ($N$ halved/doubled), (ii) vol-of-vol and correlation sweeps, and (iii) OOD drift/vol scales at inference. The Signature–RDE solver maintains monotone error decay with $N$ and degrades smoothly under parameter shifts (continuous-time hidden dynamics). Under stress vol-of-vol, Malliavin stabilization and moderate $K$ prevent gradient explosions.

\section{Experimental Details and Reproducibility Card}

We conduct all experiments on a single NVIDIA A100 (40\,GB) using PyTorch~2.x with \texttt{torchdiffeq}/\texttt{torchcde} and a signature library, fixing deterministic seeds $\{13,31,71,97,123\}$. Tasks comprise \textbf{T1 (Asian basket)} in dimensions $d\in\{20,50,100,200\}$ under Black--Scholes dynamics with the drift/vol/correlation specified in the main text and a time–average payoff; \textbf{T2 (Barrier)} with $d{=}50$ and an absorbing knock–out boundary; and \textbf{T3 (Portfolio control)} with $d{=}10$ under stochastic volatility and a risk–sensitive utility, where an HJB residual term is enabled when indicated. Discretization uses a uniform grid with $N$ steps, Euler--Maruyama for the state process $X$, and mini‐batch Monte Carlo; the grid, simulation, and cost setup are identical across all methods to ensure comparability. Architectures for our approach use signature depth $m\in\{2,3,4\}$, RDE hidden width $p\in\{64,128,192\}$, and window size $K\in\{8,12,16\}$; decoders $g_Y,g_Z,g_{\Gamma}$ are compact MLP heads, with a symmetry projection on $\Gamma$ and a mild spectral penalty. Training employs Adam/AdamW with cosine decay; we warm‐start without the HJB residual weight $\lambda_2$ and enable it after 20\% of steps, and we apply a CVaR tilt whose parameters $(q,\eta)$ are ramped linearly from $(0.95,0.5)$ to $(0.99,1.5)$ in late training. When Malliavin targets are enabled, we use antithetic sampling and pathwise weights for $Z$, while for $\Gamma$ we employ symmetricized finite‐difference control variates; all targets are computed under the same discretization to avoid bias. For accountability and measurement, we report wall–clock time per epoch (s), peak memory in GB (via the PyTorch profiler), parameter count in millions, and the exact commit hashes used. Fairness is enforced by giving all methods the same SDE simulation, time grids, cost settings, and matched parameter/time budgets; baselines are tuned via grid search following \cite{Hure2020DBDP,Beck2019DeepSplitting}.

\begin{figure}[t]
\centering
\includegraphics[width=\linewidth]{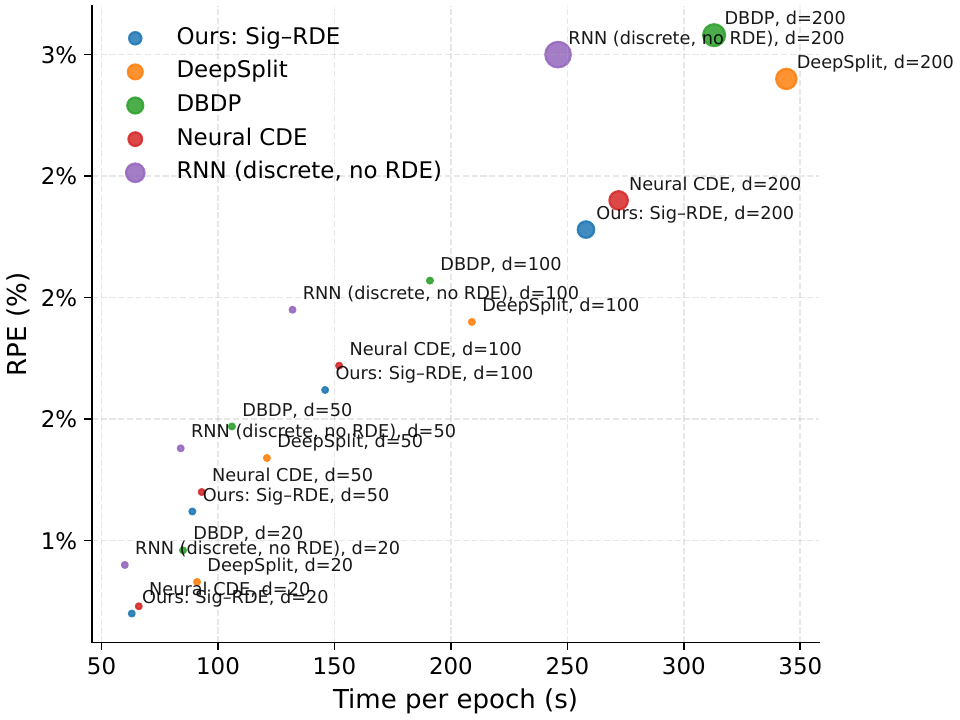}
\caption{Performance–diagnostics diagram (T1/T3): RPE vs.\ time/epoch with NaN contours (log-scale).}
\label{fig:perfdiag}
\end{figure}

\section{Discussion and Limitations}

\textbf{What the results suggest.}
Across path-dependent pricing (T1/T2) and portfolio control (T3), the Signature–RDE backbone improves both average accuracy and \emph{tail calibration} (see Figure~\ref{fig:perfdiag}). 
We read this as evidence that (i) truncated log-signatures capture long-range path information more economically than ad-hoc rolling features or discrete RNNs \cite{Feng2023DeepSignature}, and (ii) continuous-time hidden dynamics with adjoint memory mitigate exploding/vanishing gradients on long horizons \cite{Morrill2021NRDE,Fang2023NeuralRDEPPDE}. 
In fully nonlinear control, coupling with a 2BSDE head reduces variance and stabilizes second-order quantities \cite{Beck2017Deep2BSDE,PhamWarin2020FullyNonlinear}.

\textbf{When (not) to use signatures.}
Deeper truncations ($m\!>\!4$) offer diminishing returns in RPE but can still help $\mathrm{CVaR}_q$ when the payoff depends strongly on path extremes (e.g., barriers). For weakly path-dependent tasks, Markovian encoders (no-Sig ablation) may suffice and are slightly cheaper.

\textbf{Choice of RDE vs.\ discrete RNNs.}
Discrete RNNs are competitive at short horizons and small $d$; our gains increase with horizon and dimension, where continuous-time propagation and step-size control matter. The log-ODE/CDE integrator order trades accuracy for speed; our defaults target stable medium-order schemes.

\textbf{Risk-sensitive objectives.}
The CVaR-tilted terminal loss (\S3.3) consistently improves left tails without harming mean accuracy when $\eta\!\in\![1,3]$ and $q\!\in\![0.95,0.99]$. Very aggressive tilts can over-fit rare trajectories; we therefore warm-start without the tilt and enable it after terminal error plateaus.

\textbf{Second-order ($\Gamma$) estimation.}
Direct $\Gamma$ heads are flexible but noisy; Malliavin/antithetic targets reduce variance and NaNs, especially under high vol-of-vol. Autograd Hessians remain useful as diagnostics but can be brittle on long grids. A hybrid target (supervised $\hat{\Gamma}$ + physics residual) works best in our HJB runs.

\textbf{Limitations.}
(i) \emph{Truncation bias:} finite log-signature depth introduces bias for highly irregular paths; adaptive depth or learned truncation could help. 
(ii) \emph{Integrator sensitivity:} low-order solvers speed training but can under-resolve stiff residuals; implicit or adaptive solvers would be beneficial at higher cost. 
(iii) \emph{Boundaries \& constraints:} reflecting/absorbing boundaries (e.g., barriers with rebates) can degrade gradient signals; coupling with penalty methods or boundary local time terms is future work.
(iv) \emph{Model risk:} misspecified drift/vol dynamics propagate to BSDE targets; robust training (e.g., distributional shifts) remains open.
(v) \emph{Theory gap:} our consistency sketch leverages universal approximation of signatures/RDEs, but finite-sample error bounds for PPDEs with risk-tilted losses are not yet sharp.

\textbf{Relation to the two-way dialogue.}
We instantiate both directions of the workshop theme:
\emph{Finance $\rightarrow$ AI}—functional Itô/PPDE and Malliavin calculus shape the representation and variance-reduced estimators, improving interpretability and robustness \cite{Dupire2009,ContFournié2013,Ekren2011Viscosity};
and \emph{AI $\rightarrow$ Finance}—deep sequence-to-path models extend solvability of path-dependent PPDE/BSDE problems to $d\!\gg\!1$ with controllable memory/runtime via truncation depth and local windows, building on scalable BSDE training paradigms \cite{Han2018PNAS,Hure2020DBDP,Beck2019DeepSplitting,Beck2021DeepSplittingSIAM}.
In line with the workshop’s focus on stochastic control/mean-field viewpoints and risk-sensitive optimization, our CVaR tilt and 2BSDE/HJB structure align robustness with tail risk.

\section{Conclusion}

We introduced a focused deep BSDE solver for path-dependent finance that pairs \textbf{truncated log-signatures} with a \textbf{Neural RDE} backbone and, for fully nonlinear control, a \textbf{2BSDE} head. Across Asian path-dependent pricing (50D--200D) and stochastic-volatility portfolio control, the solver achieves lower errors and tighter tail calibration at comparable parameter budgets. At \(d{=}200\), \(\mathrm{CVaR}_{0.99}\) drops to \(9.8\%\) (vs.\ \(12.0\text{--}13.1\%\)) alongside smaller HJB residuals, indicating improved second-order consistency at scale. The key levers—signature depth, RDE width, local window length, and Malliavin targets—offer a practical recipe for trading accuracy, variance, and runtime: depth governs path-information capacity; width and the second-order head balance bias vs.\ variance; the local window \(K\) trades stability against wall-clock; and Malliavin targets reduce gradient noise without inflating bias.

\textbf{Outlook.} Natural extensions include (i) adaptive signature depth with error-controlled integrators; (ii) mean-field control and McKean--Vlasov couplings; (iii) diffusion-model priors for path augmentation; (iv) distributionally robust drivers for model risk; and (v) theory for finite-sample error bounds and stability of CVaR-tilted objectives in PPDE/2BSDE settings. We will release seeds, configs, and logs for full reproducibility and invite the community to stress-test the design across broader PPDE/HJB benchmarks.

\bibliographystyle{ACM-Reference-Format}
\bibliography{references_acm}

\end{document}